Selective Area Superconductor Epitaxy to Ballistic Semiconductor Nanowires

Stephen T. Gill[1*], Jeff Damasco[1], Blanka. E. Janicek[2], Malcolm S. Durkin[1], Vincent Humbert[1], Sasa Gazibegovic[3,4], Diana Car[3,4], Erik P. A. M. Bakkers[3,4], Pinshane Y. Huang[2], and Nadya Mason[1*]

1. Department of Physics, University of Illinois at Urbana-Champaign, Urbana, Illinois 61801, USA
2. Department of Materials Science and Engineering, University of Illinois at Urbana-Champaign, Urbana, Illinois 61801, USA
3. QuTech and Kavli Institute of NanoScience, Delft University of Technology, 2600 GA Delft, the Netherlands
4 Department of Applied Physics, Eindhoven University of Technology, 5600 MB Eindhoven, the Netherlands

*Corresponding Authors: gill15@illinois.edu and nadya@illinois.edu

**Abstract**

Semiconductor nanowires such as InAs and InSb are promising materials for studying Majorana zero-modes and demonstrating non-Abelian particle exchange relevant for topological quantum computing. While evidence for Majorana bound states in nanowires has been shown, the majority of these experiments are marked by significant disorder. In particular, the interfacial inhomogeneity between the superconductor and nanowire is strongly believed to be the main culprit for disorder and the resulting "soft superconducting gap" ubiquitous in tunneling studies of hybrid semiconductor-superconductor systems. Additionally, a lack of ballistic transport in nanowire systems can create bound states that mimic Majorana signatures. We resolve these problems through the development of selective-area epitaxy of Al to InSb nanowires, a technique applicable to other nanowires and superconductors. Epitaxial InSb-Al devices generically possess a hard superconducting gap and demonstrate ballistic 1D superconductivity and near perfect transmission of supercurrents in the single mode regime, requisites for engineering and controlling 1D topological superconductivity. Additionally, we demonstrate that epitaxial InSb-Al superconducting island devices, the building blocks for Majorana based quantum computing applications, prepared using selective area epitaxy can achieve micron scale ballistic 1D transport. Our results pave the way for the development of networks of ballistic superconducting electronics for quantum device applications.

**Keywords**: epitaxy, ballistic transport, nanowires, supercondivity, indium antiminoide, aluminum

**Text**

Semiconductor nanowire (NW) systems have captured great attention because of their potential use in novel quantum devices[1-4]. Recently, topological superconductivity harboring Majorana bound states has become a global pursuit in semiconductor nanowires[5,6]. While preliminary experimental work demonstrated signature zero-bias conductance peaks[7-10], the devices also demonstrated significant disorder. One notable feature was significant subgap conductance in the tunneling spectroscopy of the induced superconducting gap in the nanowire[7-10]. The so called soft-gap, which is believed to result from an inhomogenous semiconductor-superconductor interface[11], has been resolved using molecular beam epitaxy (MBE) techniques to grow epitaxial Al on InAs[12] and InSb[13] nanowires. However, such techniques are costly and not widely available to the community. In contrast, *ex-situ* processing, where metallization and wire fabrication occur in separate steps, can produce uniform interfaces and a hard superconducting at zero-magnetic field[14-15]. The growth of superconductors with suitable critical magnetic fields using conventional deposition techniques has been limited to highly disordered NbTiN[16] to date[15]. Besides disorder, NbTiN has the drawback of the development of a soft gap at finite magnetic field from vortex entry, which prevents the application to topological superconducting devices[15,17]. In contrast, nanowires with thin aluminum shells maintain robust superconductivity to roughly 2 Tesla,[18] a result of the well-known property that thin Al (thickness ≤ 10 nm) can survive in large magnetic fields.[19] Therefore, replacing NbTiN with epitaxial Al film of thickness less than 10 nm using conventional deposition techniques can provide a significant breakthrough for accessing and scaling up of topological superconductity in nanowires and nanowire networks[13,20].

In this work, we present selective area epitaxy of Al to InSb nanowires and demonstrate the high-quality electron transport enabled by epitaxial interfacing. The facets of the wire are kept sharply intact after removing the oxide using sulfur-based etching[21] while providing a low-roughness surface for epitaxial growth, as shown in Figure 1A. By engineering a clean surface without damaging the InSb crystal at the surface, Al grows epitaxially to InSb for low temperature e-beam deposition in ultrahigh vacuum (UHV) conditions, with a highly uniform interface from another device shown in Figure 1B. The Al generically induces a hard superconducting gap in the nanowire, as shown in Figure 1C, also confirming the low disorder interfacing. We also report on low temperature transport measurements for epitaxial InSb-Al devices that demonstrate superlative transport features such as near unity transmission of Andreev reflection and micron-scale ballistic transport. These results suggest that selective area epitaxy of InSb-Al nanowires is a promising system for controlling mesoscopic superconductivity, and particularly relevant for nanowire-based superconducting quantum computation ( i.e. Andreev qubits[22] and gatemons[23]).

Figure 1. Low-disorder interface and epitaxy of InSb-Al nanowires

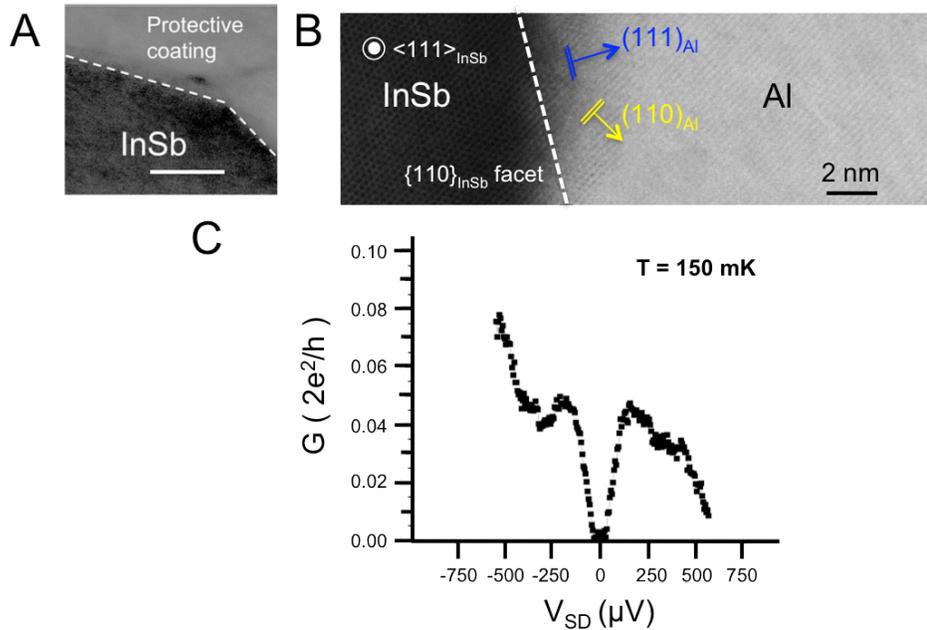

(A) Cross-sectional bright-field scanning transmission electron microscopy (BF-STEM) image of an InSb nanowire surface cleaned by a sulfur based etching and brief Ar ion mill. A metal capping layer was deposited as a protective coating before STEM imaging to protect the interface from oxidizing. A smooth interface is left on the top facet and the etching does not degrade the original faceting of the nanowire. White scale bar is 10 nm (B) Cross-sectional STEM image taken with the InSb nanowire aligned along its <111> zone axis (indicated by the dotted circle) showing the epitaxial contacting between Al and InSb following optimized etching and deposition. The white dashed line marks the InSb {110} facet. In the Al region of the STEM image, streaking is from the (110) planes of Al (double yellow lines are superimposed on these planes) growing in-plane along the nanowire's growth direction. (111) planes of Al are drawn as blue lines to indicate the out-of-plane growth orientation. Arrows indicate the normal vectors to the Al (110) and (111) lattice planes. The cross section is taken from the nanowire device shown in Figure 3A. (C) Tunneling conductance of a normal-superconducting (NS) device geometry as a function of source-drain voltage demonstrating a hard induced superconducting gap.

First, we discuss the procedure for selective area epitaxy of Al to InSb nanowires. Fabrication begins by depositing nanowires on a pre-patterned chip with a micromanipulator. See Supporting Information for additional details regarding chip preparation. A transferred wire is shown in Figure 2A. Next, we use conventional e-beam lithography for masks in all etching and deposition steps, as schematically demonstrated in 2B-C. To remove the native oxide, a key step for preparing a transparent interface, we use sulfur passivation as a first step in surface cleaning[21]. Following sulfur passivation, the sample is transferred to a load-locked UHV system where the chip is outgassed for several hours until base pressure of the load lock chamber is achieved. After outgassing the chip, a brief, low-energy Ar ion mill is performed to remove any adsorbates on the nanowire surface. As evidenced in Figures 1A-B, this processing produces a nearly disorder-free surface where there is no visible amorphous layer at the surface of the nanowire. Indeed, recent work has shown that by optimizing sulfur-based etching, low disorder InSb nanowire interfaces can be prepared for depositing superconductors and inducing a hard gap comparable to the gap "hardness" demonstrated in MBE based InAs-Al nanowires[14-15]. However, this work on ex-situ based superconductor deposition required a disordered sticking metal, such as NbTi or Ti, and there was non-epitaxial growth of the parent superconductor to achieve a hard gap. In that case it was not possible to interface the nanowire with Al, which was attributed to non-epitaxial growth[14]. Similarly, we have deposited Al at room temperature onto InSb nanowires using similar procedures as in Ref. 14 and have measured contact resistances on the order of MΩs.

We now outline the conditions required for selective area epitaxy of Al following surface preparation, which is schematically represented in Figure 2C. The growth occurs at liquid nitrogen temperature with a low background pressure of ~ 1.5-4×10$^{-10}$ torr during the cooling and before deposition. Once the sample is at liquid nitrogen temperature, high-purity Al is e-beam evaporated. In our growth procedure, Al films are deposited at a rate of ~ 0.1 A/s. The Al flux is linearly directed at the sample, as illustrated in Figure 2C-D. Scanning transmission electron microscopy (STEM) characterization of a typical device reveals areas of epitaxial interfacing, as shown in Figure 1B. The TEM image, taken where the InSb nanowire is aligned along the <111> zone axis, shows the Al (110) planes growing in plane to the top InSb {110} facet. The Al (110) planes seen along the <111> orientation are a result of Al (111) growth out of plane to the <111> nanowire growth direction. As pointed out in Krogstrup *et. al.*[12], the low temperature is critical for achieving crystalline thin films of a single grain where minimizing surface free energy is the dominant mechanism for determining the out of plane crystal orientation. For FCC metals such as Al, the lowest surface energy orientation is generally (111). Hence, our TEM analysis confirms we are growing epitaxial Al films where surface free energy minimization is the strongest thermodynamic driver in dictating the films out of plane crystallinity. After deposition and a several-hour warm up to room temperature, a standard liftoff in acetone is used to remove resist, and, as shown in Figure 2E, a smooth film of Al is left on the nanowire. Our growth conditions reproducibly result in the growth of a continuous, smooth Al shell onto InSb. In contrast to the elevated temperature growth of epitaxial Al on InAs[24], we do not observe out of plane variations, which would indicate polycrystalline growth along the nanowire growth direction. As discussed above, the lack of out plane variations is a result of the low temperature growth promoting a large area grain along the nanowire having a single out of plane crystal orientation.

Figure 2. Schematic of processing for patterned epitaxy

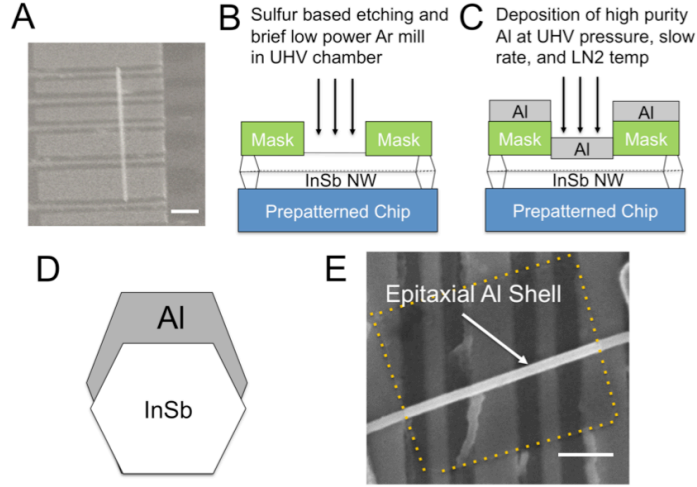

(A) SEM image of an InSb nanowire transferred to a pre-patterned chip with local gate electrodes. Scale bar is 750 nm
(B) Schematic of surface cleaning for preparing clean InSb surfaces. Nanowires are patterned with masks using e-beam lithography. The mask is used to selectively etch and clean the nanowire using sulfur passivation and ion milling.
(C) Schematic of deposition conditions required for epitaxial growth. (D) Illustration of the general growth of Al onto InSb in our system. The deposition source is located 1 m away from the sample, so the growth is highly directional and Al grows dominantly on the top facet. (E) SEM image of a completed deposition where a nominal 8 nm epitaxial Al shell is left on the nanowire. The yellow dashed lines outline the mask which was used to etch and selectively grow Al on the nanowire. Scale bar is 200 nm.

The surface homogeneity of the nanowire-superconductor surface and the crystalline superconducting film enables the superconductor to impart a hard superconducting density of states in the nanowire. As shown in Figure 1C, the tunneling regime of a ballistic normal-superconductor (NS) nanowire device shows strong suppression of conductance below the superconducting gap. In particular, the measured ratio of zero bias conductance ($G_{V=0}$) to normal state conductance ($G_N$) in this device reveals a suppression of $G_{V=0}/G_N < 1/50$, comparable to gap hardness in MBE based Al-InAs nanowires[25,26]. In addition to strong suppression of subgap conductance in the tunneling regime, we find that the device achieves a large zero-bias Andreev enhancement to *1.7 $G_0$* in the single mode regime of the QPC (see Supporting Information), indicating 96% transmission[14,25,26]. Further, a 2D map of conductance as a function of gate and source-drain voltage reveals the device evolves from tunneling to the single mode regime without localization features, also consistent with highly transparent leads (see Supporting Information).

We now discuss the high-quality ballistic 1D superconductivity that can be achieved in our epitaxial InSb-Al nanowires. In particular, we focus on the transparency of the epitaxial InSb-Al nanowire interface in the single mode regime, which is relevant for engineering Majorana modes[5,6]. As we have demonstrated previously, InSb nanowires interfaced with Al can be engineered to have a quantum point contact response (i.e., quantization of conductance steps), which allows gate voltage tuning to the single mode regime[27]. Figure 3A shows an Al-InSb-Al nanowire Josephson junction where quantized transport of the normal state conductance is observed, as shown in the inset of Fig. 3B. (See Supporting Information for detail regarding the device fabrication). The main part of Figure 3B shows that the first quantized conductance plateau, at ~ $2e^2/h$, is enhanced in the superconducting regime to ~ $4e^2/h$. This enhancement is caused by Andreev reflection, the process of generating superconducting proximity effect by an electron reflecting as a hole at the nanowire-superconductor interface[28]. For pristine superconducting-normal interfaces, Andreev reflection can enhance conductance by a factor of 2[28]. In SNS systems, the crossover from Andreev to normal conductance occurs at $eV_{SD} = 2\Delta_0$, where $\Delta_0$ is the induced gap in the proximitized region and the factor of 2 accounts for two S-N interfaces in series. For the device in Fig. 3, we observe a resonant superconducting behavior where the conductance closely reaches the theoretical limit of $4e^2/h$ at $eV_{SD} = 2\Delta_0$, indicating a near unity Andreev reflection probability at the interface, when the normal conductance is tuned to the middle of the single mode conductance plateau. Figure 3C shows the bias dependence on resonance where a strong Andreev conductance enhancement to $4e^2/h$ is observed in the crossover from normal to Andreev conductance, $V \leq 2\Delta_0$, where $\Delta_0 = 125$ μeV is the induced gap in the proximitized region from the Ti/Al leads. While the presence of multiple Andreev resonances is typically presented to imply a transparent interface, we note that the disappearance of these resonances is expected for the zero-scattering and low temperature limit of multiple Andreev reflection theory[29-31]. Additionally, enhanced conductance above $2e^2/h$ for small voltages above $eV_{SD} = 2\Delta_0$ is from excess current[28], which we have observed before in superconducting InSb QPCs[27]. Note the small zero bias peak imposed on the broader enhanced Andreev conductance is due to a supercurrent. An 8 μV excitation was used in the differential conduction measurement which suppresses the magnitude of the zero-bias peak from the supercurrent. When the device is further opened with the backgate to allow additional modes, Andreev transport becomes sharply suppressed. Similar resonant Andreev enhancement behavior has been observed in InSb nanowires[14] and in quantum point contacts (QPC) formed in epitaxial InAs-Al 2DEGs[32].

Figure 3. Andreev doubling and transparent transmission of Cooper pairs through an epitaxial Al-InSb-Al junction

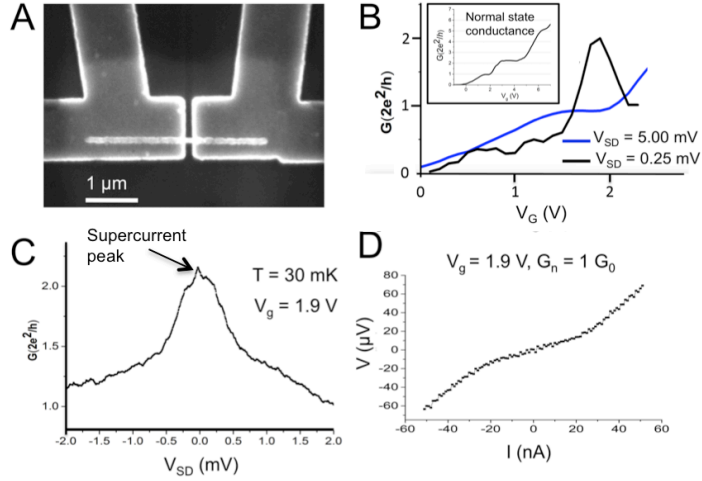

(A) SEM image of an Al-InSb Josephson junction whose epitaxial interfacing is shown in figure 1B. (B) Differential conductance as a function of gate voltage $V_G$ at $V_{SD}$ = 2$\Delta_0$ = 250 µV, i.e. single Andreev reflection conductance, (black line) and at $V_{SD}$ = 5 mV >> 2$\Delta_0$ (blue line) where the device is normal. On a resonance value of $V_G$ = 1.9 V, the Andreev conductance nearly doubles from the normal state value, as expected for a pristine interface. Inset shows normal conductance quantization for the device. (C.) Differential conductance as a function of source-drain voltage of the device on the resonant gate voltage of of $V_G$ = 1.9 V. For $V_{SD}$ < 2$\Delta_0$, the device conductance rises above twice the normal state conductance, signaling SNS behavior. Small offset of the peak from $V_{SD}$ = 0 mV is from an instrumen offset. (D) IV behavior of the device on resonance showing a switching current of roughly 24 nA.

The transparent InSb-Al interface is also evident from the switching current measured when the device is tuned to resonant Andreev transport. Given the resonant Andreev behavior, we observe switching current resonances rather than plateaus, consistent with the correlation of Andreev transport to switching currents (see Supporting Information). Figure 2D shows a switching current of $I_S \approx$ 24 nA when the device is tuned to the gate voltage where Andreev doubling appears. We note that the residual resistance (less than 1 kΩ and more than a factor of 10 smaller resistance than the normal state resistance) observed below the switching current, $I_S$, is from phase diffusion in a small, unshunted junction[33,34] and is commonly observed when the normal-state resistance is large[35,36]. The switching current for a perfect S-QPC-S junction at unity transmission is given by $I_N = 2\pi N e \Delta_0 / h$, where $N$ is the number of modes and $\Delta$ is the induced superconducting gap[37]. For a gap of 125 µeV, the maximum supercurrent for a single mode weak link would be 30 nanoamps. Accounting for the deviation from an ideal junction by finite reflections[38], we extract a transmission of 96%. Consideration of thermal suppression would lead to an even higher transmission value[33,38,39], consistent with having achieved a nearly pristine epitaxial interface. We have measured similar behavior in multiple other devices, indicating that pristine interfaces can be achieved reproducibly and that epitaxial Al-InSb NW Josephson junctions can demonstrate nearly ideal mesoscopic superconductivity.

Next, we demonstrate that not only is transport ballistic in the quantum point contact formed in the bare nanowire between contacts, but that the epitaxial Al-InSb NW segment is also a ballistic quantum wire. One of the major developments enabled by the growth of epitaxial Al-InAs nanowires was the development of gate tunable superconducting islands for studying 1D topological superconductivity, i.e. the so-called Majorana island geometry[40,41]. We focus on the behavior of epitaxial InSb-Al nanowires in the island geometry, with a device and schematic shown in Figure 4A and 4B. We note that the devices tested had a single plunger gate under the superconducting shell. Thin gates underneath the bare nanowire form quantum point contacts coupling the epitaxial InSb-Al nanowire to the leads. For two quantum point contacts in series to a ballistic reservoir, the resistance is modulated as $R_{Tot} = Max(R_{QPC1}, R_{QPC2})$[42]. In contrast, a diffusive reservoir's conduction would be modulated by Ohmic addition, ($R_{Tot} = R_{QPC1} + R_{QPC2} + R_{inelastic}$)[42]. The plot in Figure 4C is of a schematic of the expected behavior for the conductance for two quantum point contacts in series with a ballistic reservoir. The plot on the right in Figure 4D shows the response of a 600 nm long epitaxial InSb-Al quantum wire (5 nm epitaxial film of aluminum) coupled to two 200 nm long quantum point contacts. For this measurement the plunger gate is grounded and a finite voltage is applied to drive the device normal. We see that the conductance in this device shows behavior corresponding to quantum addition where the conductance is dominated by the plateau behavior of the QPCs. As shown in 4E, a cut from the plot shows a robust plateau at $dI/dV = 2e^2/h$, with small mesoscopic fluctuations imposed, indicating ballistic, single mode transport. The quantization quality is comparable to what is seen in long quantum wires[43,44]. This transport behavior demonstrates that there is ballistic transport through the Al-InSb NW portion, and that there is quantum point contact coupling of the superconducting NW to the leads. Hence, electrons are travelling more than a micron without scattering in these structures. Even more importantly, the data suggests the NW has been tuned into a regime where there is most-likely a single mode flowing in the semiconductor-superconductor NW, realizing the required single-mode quantum wire regime to engineer Majorana zero-modes and observe helical transport[6].

Figure 4. Micron-scale ballistic transport and engineered confinement in an epitaxial InSb-Al device

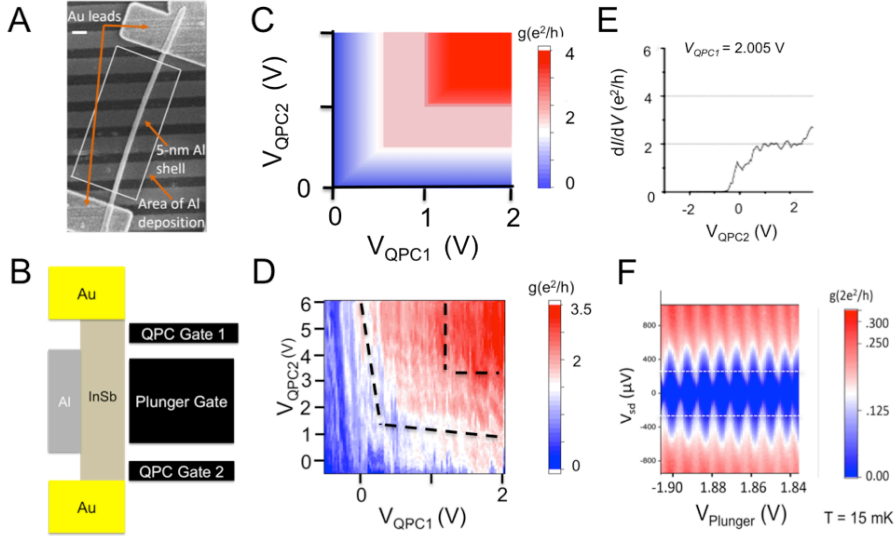

(A) SEM image of a representative epitaxial InSb-Al island device. An Al island of thickness below 10 nm is deposited selectively between two gold leads. Local gates tune the QPC in the bare regions between the leads and island, and a plunger gate tunes the chemical potential under the proximitized InSb nanowire. White scale bar is 200 nm (B) Schematic of the epitaxial InSb-Al island device. Two gates acting as QPCs in the bare constrictions between Au leads control the device from Coulomb blockaded to ballistic quantum wire regimes. A plunger gate is used to tune the chemical potential in the epitaxial InSb-Al nanowire. (C) Left, schematic of the conductance as a function of arbitrary gate voltage on two series QPCs when transport through a reservoir is ballistic according to Landauer's formula[42]. The conductance is dominated by plateaus indicating how many modes are transmitting. (D) conductance as a function of two series QPCs in a 600 nm island device (Al thickness of 5 nm) when the plunger is grounded. As highlighted by the dashed lines, conductance is dominated by the quantized conductance plateau for single mode transmission, indicating quantum addition and ballistic transport of a single mode through the InSb-Al island. Note, a 1 mV bias was applied to suppress superconductivity (E) Cut of the conductance map in (D) for the gate voltage on QPC 1 held at $V_{QPC1}$ = 2.005 V. A robust quantized conductance plateau is seen with minor fluctuations imposed, indicating ballistic transmission of a single mode through the InSb-Al island. (F) Map of differential conductance as a function of source-drain voltage and plunger voltage showing Coulomb blockade. Both QPC gates are held at -0.45 V to deplete the QPC. The superconducting gap is $\Delta_0$ = 250 μeV and the charging energy is 350 μeV.

In the tunnel-coupled limit, the device operates as an isolated superconducting island in the Coulomb blockade regime. The epitaxy technique enables the use of very thin films – we use < 10 nm Al for islands – that can be strongly gated to tune the proximitized nanowire. Figure 3D shows the Coulomb blockade behavior of the InSb-Al NW from Figure 3C when the QPC gates are tuned to pinch-off. The plunger gate modulates the single electron Coulomb blockade with a periodicity of roughly 7 mV and with an extracted charging energy of 350 μeV, giving a lever arm of η = .05 eV/V, indicating strong coupling of the plunger to the proximitized nanowire. This result indicates the extra metal deposited in the epitaxial contacting to the nanowire does not significantly screen the gate. We note that connecting the additional metal deposited in the epitaxial contacting of the nanowire to the plunger can enhance the plunger coupling (*i.e.* decrease the charging energy). This is relevant for tuning the superconducting island to a regime where $\Delta_0 > E_C$ in order to observe 2e periodic Coulomb blockade and measure the topological transition to 1e periodic Majorana quasiparticle teleportation[41]. Our ability to strongly tune the chemical potential of the superconducting quantum wire will allow us in future experiments to explore the topological phase diagram of InSb-Al nanowires and the signature of exponential protection in the Majorana island geometry[41].

While our epitaxial growth procedure is applicable to other nanowire systems (*e.g.* InAs nanowires), epitaxial InSb-Al NWs offer several advantages for studying topological superconductivity over other NW systems. For example, the g factor of InSb is roughly a factor of 5-10 larger than InAs NWs[15,45-47], significantly bringing down the required magnetic field to enter the topological regime. Additionally, InAs nanowires, including the epitaxial Al-InAs system, are prone to developing unintentional quantum confined regions from surface sensitivity in InAs[18,48]. For the epitaxial InAs-Al system, device fabrication requires a harsh etch to remove unwanted Al, which results in random quantum dot formation[18]. Indeed, signatures of Majorana in these devices show zero-energy bound states with spectral weights far below the expected quantized values, and broadening inconsistent with the Majorana interpretation, despite the improved proximity superconductivity[40]. These results suggest that disorder is still playing a large role in the readout of Majorana zero-modes and that quantum control in the system is lacking. However, InSb nanowires devices can be *ex-situ* processed to give a full yield of devices showing ballistic 1D transport, as evidenced by robust quantized conductance at zero-magnetic field and miliKelvin temperatures[15,27,49]. We note that the InSb-Al system has never yielded unintentional Coulomb blockaded, quantum dot-like transport. In particular, we have measured Andreev enhancement and switching current correlation to the normal state conductance quantization in approximately 20 devices. We note that the observation of approximate Andreev doubling and switching currents approaching the quantized limit has only been observed in several devices, as this requires high quality normal state conductance quantization and hence extremely clean wires. We believe such high quality ballistic device should show rich behavior in a magnetic field. In particular, future measurements will test the behavior of the induced gap of thin Al shell devices in magnetic field in order to perform quantum point contact spectroscopy experiments of Majorana states[50].

In conclusion, we have presented selective area superconductor epitaxy to semiconductor nanowires. In particular, we have focused this technique on InSb nanowires to develop superconducting quantum wires. We have demonstrated that epitaxial growth of Al following sulfur-based removal of the native oxide results in a nearly ideal superconducting interface. As a result of this clean interface, we realize near unity Andreev reflection at the InSb-Al interface, which gives large conductance enhancements from Andreev reflection and nearly unity transmission of supercurrents in the single mode regime. In addition, we have shown that this method is amenable to developing ballistic superconducting island devices with QPC coupling to the leads. Our quantum point contact control in InSb-Al nanowires should enable proper determination of the topological transition[51] and further establish epitaxial InSb-Al nanowires as a promising platform for studying and manipulating Majorana modes.

**Note**

The authors declare no conflict of interest.

## Supporting Information Available

The Supporting Material is available free of charge via the Internet at http://pubs.acs.org/ Contained are details of the fabrication, the measurement setup, TEM imaging, and the treatment of contact and series resistance along with extended data and discussion for the devices in Figures 1, 3, and 4.

## Acknowledgments

S.T.G, J.D., M.S.D, V.H., and N.M. were supported by the Office of Naval Research Grant No. N0014-16-1-2270. B.E.J. and P.Y.H. were supported by the Office of Naval Research Grant No. N00014-16-1-2436. S.G, D.C., and E.P.A.M. B. were partly supported by the European Research Council and The Netherlands Organization for Scientific Research. S.T.G acknowledges support from an NSF Graduate Research Fellowship. S.T.G, J.D., B.E.J., M.S.D., V.H., P.Y.H, and N.M. acknowledge the use of the Materials Research Lab Central Facilities at the University of Illinois for all work.